\documentclass[aps, amsmath, amssymb, aps, pre, notitlepage,%
oneside, onecolumn,groupedaddress,superscriptaddress,nofootinbib,longbibliography]{revtex4-1}

\usepackage{amsfonts, amsmath, amssymb, amsthm, xfrac}
\usepackage{enumitem}
\setlist{noitemsep,topsep=0pt,parsep=0pt,partopsep=0pt}
\usepackage{array}
\usepackage[russian,english]{babel}
\usepackage[T2A]{fontenc}
\usepackage[utf8]{inputenc}
\usepackage{graphics}
\usepackage{graphicx}
\usepackage[margin=10pt,parskip=5pt,indention=10pt, lofdepth,lotdepth]{subfig}
\usepackage{listings, xcolor}
\usepackage{bm}
\usepackage{epstopdf}
\usepackage{lineno,hyperref}
\modulolinenumbers[5]

\usepackage{epigraph}
\newtheorem{theorem}{Theorem}

\usepackage[normalem]{ulem}
\usepackage[colorinlistoftodos,prependcaption,textsize=medium]{todonotes}

\DeclareMathOperator{\LEs}{LE}

\begin{document}

\title{Numerical analysis of dynamical systems: unstable periodic orbits,
hidden transient chaotic sets, hidden attractors, and finite-time Lyapunov dimension}

\author{N. V. Kuznetsov}
\email[]{Corresponding author: nkuznetsov239@gmail.com}
\affiliation{Faculty of Mathematics and Mechanics, St. Petersburg State University,
Peterhof, St. Petersburg, Russia}
\affiliation{Department of Mathematical Information Technology,
University of Jyv\"{a}skyl\"{a}, Jyv\"{a}skyl\"{a}, Finland}
\affiliation{Institute of Problems of Mechanical Engineering RAS, Russia}
\author{T. N. Mokaev}
\affiliation{Faculty of Mathematics and Mechanics, St. Petersburg State University,
Peterhof, St. Petersburg, Russia}

\begin{abstract}
In this article, on the example of the known low-order dynamical models, namely
Lorenz, R\"{o}ssler and Vallis systems,
the difficulties of reliable numerical analysis of chaotic dynamical systems
are discussed.
For the Lorenz system, the problems of existence of hidden chaotic attractors and
hidden transient chaotic sets and their numerical investigation
are considered.
The problems of the numerical characterization of a chaotic attractor
by calculating finite-time time Lyapunov exponents and finite-time Lyapunov dimension
along one trajectory are demonstrated using
the example of computing unstable periodic orbits in the R\"{o}ssler system.
Using the example of the Vallis system describing the El~Nin\~{o}-Southern Oscillation
it is demonstrated an analytical approach for localization of self-excited and hidden attractors,
which allows to obtain the exact formulas or estimates
of their Lyapunov dimensions.
\end{abstract}

\maketitle

\section{Introduction}

History of the turbulence phenomena study
is associated with the consideration of various models,
which include the Navier-Stokes equations,
their Galerkin approximations, and the development of the theory of chaos
\cite{Landau-1944,Hopf-1948,RuelleT-1971,Smale-1967}.
Here let us note the significant results by
D.~Ruelle, F.~Takens \cite{RuelleT-1971}, and S.~Smale \cite{Smale-1967},
who proposed a chaotic attractor as a mathematical prototype describing
the onset of turbulence, and by O.~Ladyzhenskaya, who
studied the case when the two-dimensional Navier-Stokes equation
generates a dynamical system and proved the
finite dimensionality of its attractor \cite{Ladyzhenskaya-1982}.
The first vivid example of chaotic attractor
in a hydrodynamic system was obtained
by E.~Lorenz \cite{Lorenz-1963}.
Using the Galerkin method
he derived a crude three-dimensional mathematical model
for Rayleigh-B\'{e}nard convective flow, which has
the following form:
\begin{equation}\label{sys:lorenz}
\begin{cases}
 \dot{x} =  - \sigma(x-y),\\
 \dot{y} = r x - y - x z, \\
 \dot{z} = - b z + x y,
\end{cases}
\end{equation}
where $r$, $\sigma$, $b$ are positive parameters.
For $0 < r < 1$, there is one globally asymptotically stable equilibrium $S_0 = (0,0,0)$.
For $r > 1$, equilibrium $S_0$ is a saddle, and a pair of symmetric equilibria
$S_\pm = (\pm\sqrt{b(r-1)},\pm\sqrt{b(r-1)},r-1)$ appears.

For the parameters $r = 28$, $\sigma = 10$, $b = 8/3$ in system \eqref{sys:lorenz}
E.~Lorenz numerically found a chaotic attractor in the model.
In general, for numerical localization of attractor, it is necessary
to explore its basin of attraction and choose an initial point in it.
If for a particular attractor its basin of attraction is connected with the
unstable manifold of unstable equilibrium, then the localization
procedure is quite simple.
From this perspective, the following classification of attractors
is suggested \cite{LeonovKV-2011-PLA,LeonovK-2013-IJBC,LeonovKM-2015-EPJST,Kuznetsov-2016}:
an attractor is called a \emph{self-excited attractor}
if its basin of attraction intersects an arbitrarily small open neighborhood of an equilibrium;
otherwise, it is called a \emph{hidden attractor}.
Numerical localization of hidden attractors is much
more challenging and requires the development of special methods.
The classical Lorenz attractor is a self-excited one with respect
to all equilibria $S_0$, $S_\pm$, and it is an open question~\cite[p.~14]{Kuznetsov-2016}
whether for some parameters there exists a hidden Lorenz attractor.
This question is related to the ''chaotic'' generalization~\cite{LeonovK-2015-AMC}
of the second part of Hilbert's 16th problem \emph{on the number and mutual
disposition of attractors and repellers in the chaotic multidimensional dynamical systems and,
in particular, their dependence on the degree of polynomials in the model};
see corresponding discussion, e.g. in \cite{SprottJKK-2017,ZhangChen-2017}.
There are a number of physical dynamical models which possess hidden chaotic attractors,
e.g, the Rabinovich system
(describes the interaction of plasma waves)~\cite{KuznetsovLMPS-2018,ChenKLM-2017-IJBC},
the Glukhovsky-Dolghansky system
(describes convective fluid motion in a rotating cavity)~\cite{LeonovKM-2015-CNSNS,LeonovKM-2015-EPJST}
and others~\cite{DancaKC-2016,DudkowskiJKKLP-2016,GarashchukSK-2018-HA,GarashchukSK-2018-RCD}.

We note that the Lorenz system \eqref{sys:lorenz} with parameters
$r = 28$, $\sigma = 10$, $b = 8/3$ is dissipative in the sense of Levinson,
and for any initial data (except for equilibria) the trajectory tends to the attractor.
Thus, system \eqref{sys:lorenz} generates a \emph{dynamical system}
$\big(\{\varphi^t\}_{t\geq0}, (U\subseteq \mathbb{R}^3, ||\cdot||)\big)$.

\section{Hidden transient chaotic sets in Lorenz system}

In numerical computation of a trajectory over a finite-time interval
it is difficult to distinguish a \emph{sustained chaos} from
a \emph{transient chaos}
(a transient chaotic set in the phase space, which can nevertheless persist for a long time)
\cite{GrebogiOY-1983,LaiT-2011}, thus it is reasonable to give
a similar classification for transient chaotic sets~\cite{DancaK-2017-CSF,ChenKLM-2017-IJBC}:
a \emph{transient chaotic set} is called a \emph{hidden transient chaotic set}
if it does not involve and attract trajectories
from a small neighborhood of equilibria;
otherwise, it is called \emph{self-excited}.
In order to distinguish an attracting chaotic set (attractor)
from a transient chaotic set in numerical experiments,
one can consider a grid of points in a small neighborhood of the set
and check the attraction of corresponding trajectories towards the set.
There one can reveal a subset of points for which the trajectories leave
the transient set.

\begin{figure}[!ht]
 \centering
 \subfloat[
 {\scriptsize $u_0\!=\!(2,2,2)$, $t \in [0,1000]$}
 ] {
 \label{fig:lorenz:attr:hidTrans:hid}
 \includegraphics[width=0.3\textwidth]{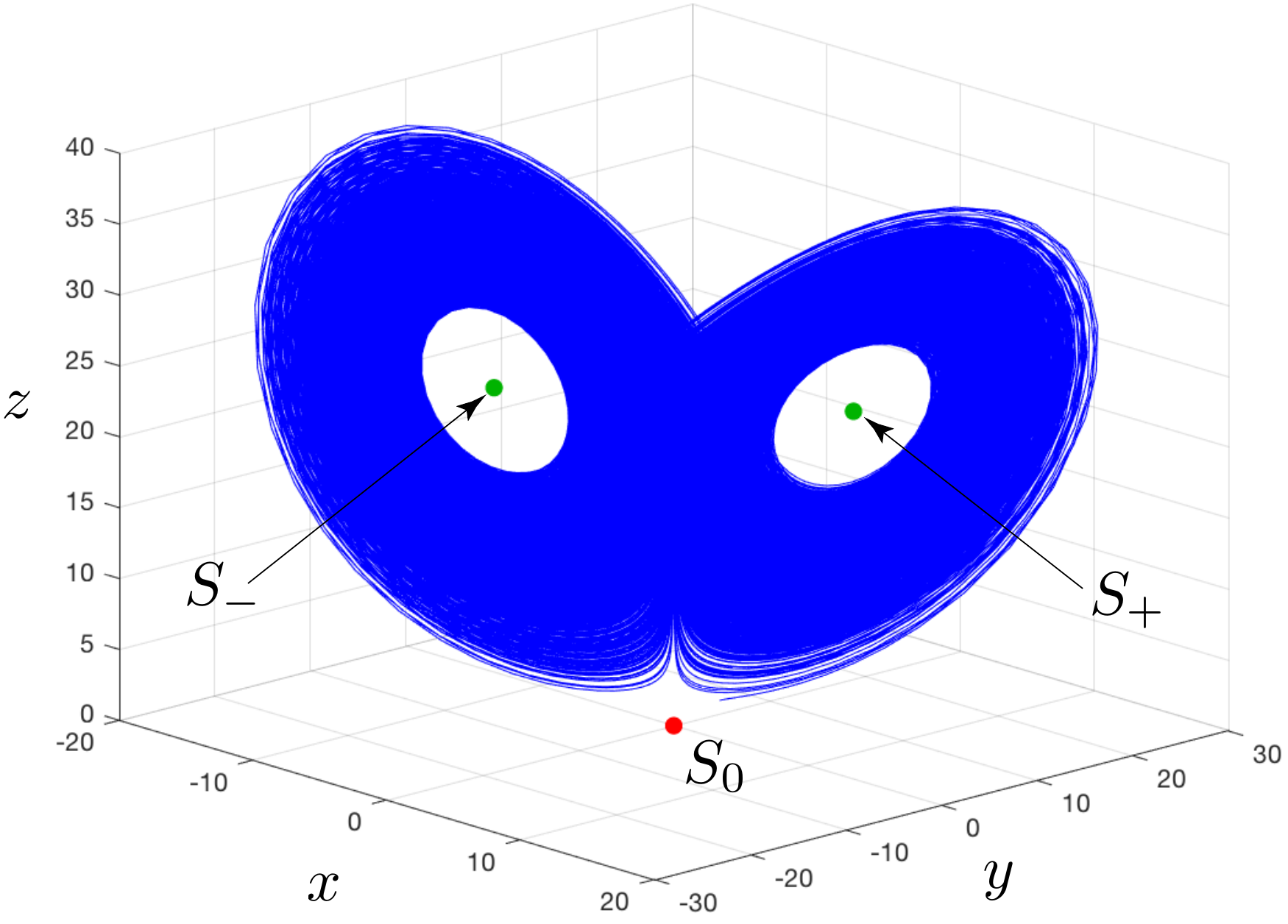}
 }~
 \subfloat[
 {\scriptsize $u_0\!=\!(2,2,2)$, $t \in [0,2.3\cdot10^{5}]$}
 ] {
 \label{fig:lorenz:attr:hidTrans:trans}
 \includegraphics[width=0.3\textwidth]{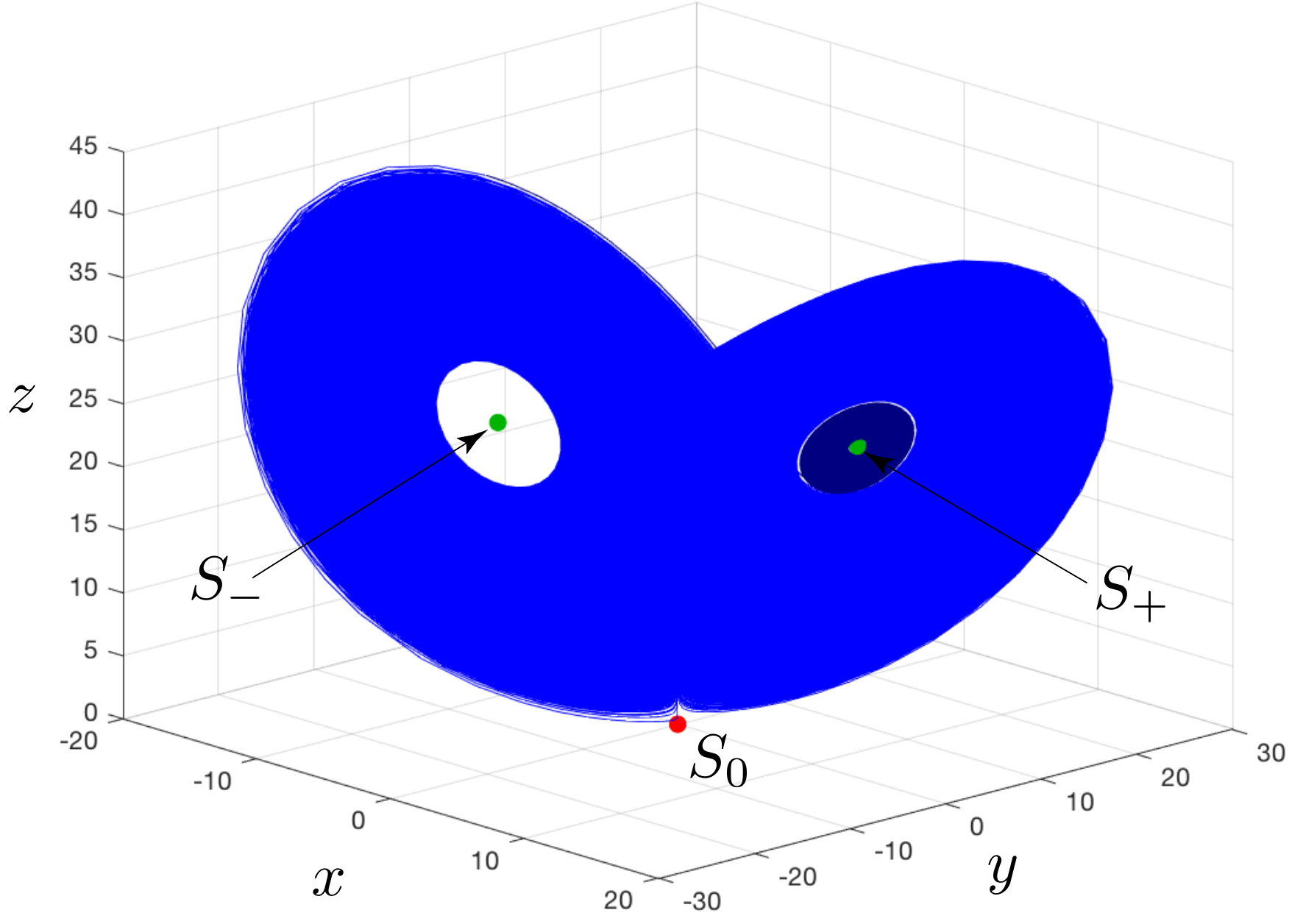}
 }
 \subfloat[{\scriptsize $u_0$ in vicinity of $S_0$}
 ] {
 \label{fig:lorenz:attr:hidTrans:sepa}
 \includegraphics[width=0.3\textwidth]{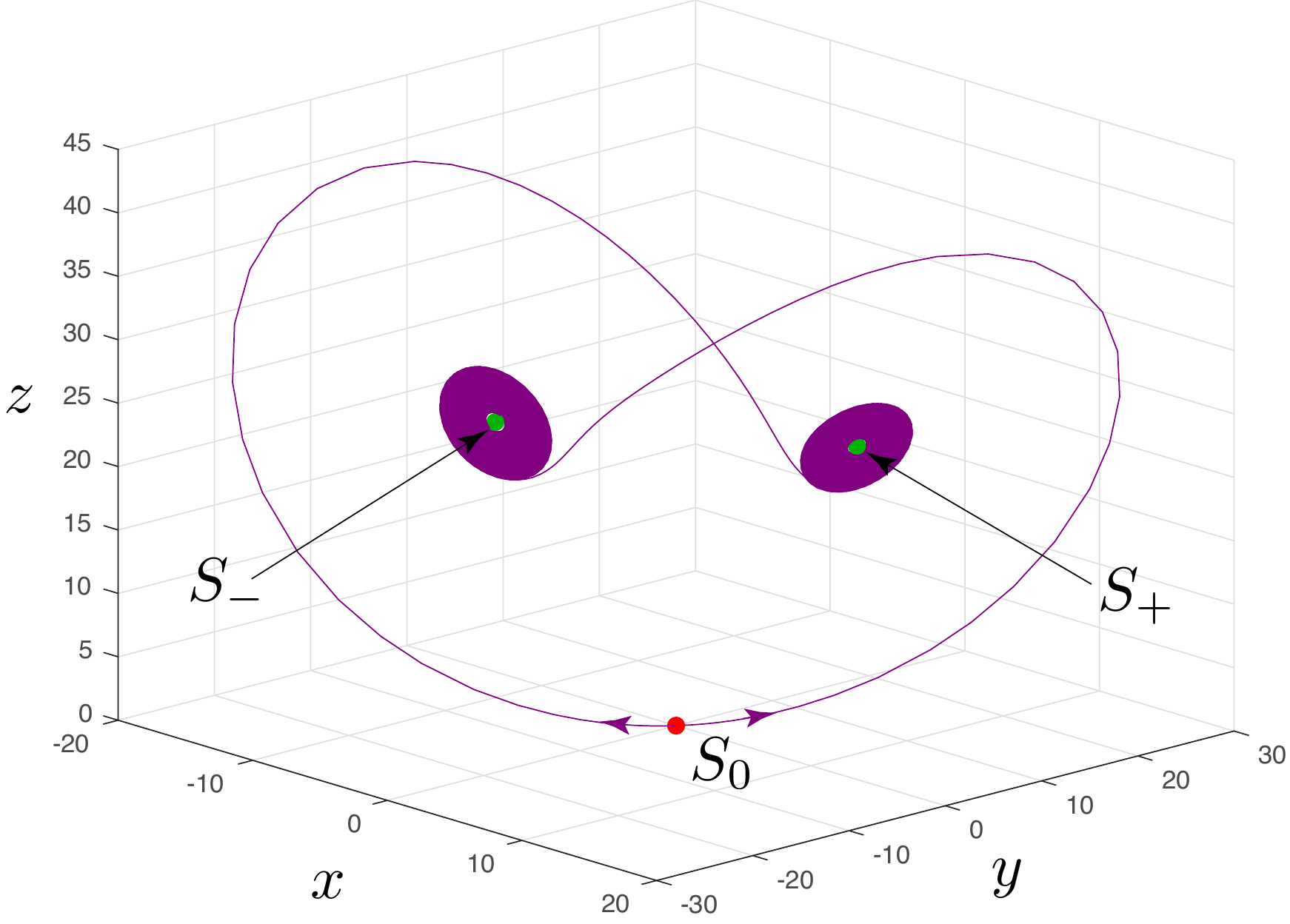}
 }
 \caption{
 Visualization of
 the hidden transient chaotic set in system~\eqref{sys:lorenz},
 $r = 24$, $\sigma = 10$, $b = \tfrac{8}{3}$.
 }
 \label{fig:lorenz:attr:hidTrans}
\end{figure}

For the Lorenz system \eqref{sys:lorenz},
suppose that $\sigma = 10$, $b = \tfrac{8}{3}$ are fixed
and $r$ varies.
If $r = 28$, then all three equilibria $S_0$, $S_\pm$ are unstable and
in the phase space there exists a self-excited attractor with respect to these equilibria.
For $r < \sigma (\tfrac{\sigma + b + 3}{\sigma - b - 1}) \approx 24.7368$,
the equilibria $S_\pm$ become stable, and
for  $24.06 < r < 24.7368$, there exists a self-excited attractor with respect to equilibrium $S_0$.
Near the point $r \approx 24.06$ it is possible
to observe a long living transient chaotic set, which is hidden
since it's basin of attraction does not intersect with the small vicinities of equilibrium $S_0$.
For example, for $r = 24$ a hidden transient chaotic set
can be visualized\footnote{
 In this work, we use MATLAB's standard procedure \texttt{ode45} with default parameters.
}~\cite{YuanYW-2017-HA}
from the initial point $(2,2,2)$ (see Fig.~\ref{fig:lorenz:attr:hidTrans}).
In~\cite{MunmuangsaenS-2018-HA}, hidden transient chaotic set
was obtained in system \eqref{sys:lorenz} with $r = 29$, $\sigma = 4$, $b = 2$.

The time of the transient process in this case depends strongly
on the choice of the initial data, which complicates the task of distinguishing
an attracting chaotic set (attractor)
from a transient chaotic set in numerical experiments.
E.g., for system \eqref{sys:lorenz} with parameters $r = 24$, $\sigma = 10$, $b = 8/3$
and for initial point $u_0 = (20,20,20)$
a transient chaotic behavior is observed\footnote{
  The time of transient chaotic behavior is often estimated approximately
  by analyzing the sign of the largest Lyapunov exponent.
  For simplicity, here we approximate the transient behavior time
  by the time of entering a small ball with the center at the points~$S_\pm$.
} on the time interval $[0,\,1.8\!\cdot\!10^{4}]$,
for initial point $u_0 = (-7,8,22)$ --- on the time interval $[0,\,7.2\!\cdot\!10^{4}]$,
for initial point $u_0 = (2,2,2)$ --- on the time interval $[0,\,2.26\!\cdot\!10^{5}]$,
and for initial point $u_0 = (0, -0.5, 0.5)$ a transient chaotic behavior
continues over a time interval of more than $[0,\,1\!\cdot\!10^{7}]$.
In order to distinguish an attracting chaotic set
from a transient chaotic set by computing trajectories on a reasonable time interval,
one can consider a grid of points in a small
neighborhood of the set and check the attraction of corresponding
trajectories towards the set.
There one can reveal a subset of points for which the trajectories leave
the transient set.

Next, on the example of the Lorenz system \eqref{sys:lorenz} we will demonstrate
difficulties in the reliable numerical computation of the
finite-time Lyapunov exponents and
finite-time Lyapunov dimension.

\section{Finite-time Lyapunov dimension of a transient chaotic set}

Consider system \eqref{sys:lorenz} with parameters
$r = 24$, $\sigma = 10$, $b = 8/3$ and integrate numerically
the trajectory with initial data
$u_{0} = (20,\,20,\,20)$.
We numerically approximate the finite-time Lyapunov exponents
and finite-time Lyapunov dimension (see corresponding definitions, e.g.
in ~\cite{Kuznetsov-2016-PLA,KuznetsovLMPS-2018}).
Integration with $t > T_1 \approx 1.8\!\cdot\!10^{4}$ leads to the collapse of the ``\emph{attractor}'',
i.e. the ``\emph{attractor}'' turns out to be a transient chaotic set.
However, on the time interval $t \in [0,~T_3\approx 507883]$
we have $\LEs_1(t, \, u_{\rm init}) > 0$
and, thus, may conclude that the behavior is chaotic,
and for the time interval $t \in [0,~T_2\approx 262954]$
we have ${d}_{\rm L}^{\rm KY}(\{\LEs_i(t, \, u_{\rm init})\}_{i=1}^3) > 2$.
This effect is due to the fact that the finite-time Lyapunov exponents and
finite-time Lyapunov dimension are the values averaged over the considered time interval.
Since the lifetime of transient chaotic process can be extremely long
and taking into account the limitations of reliable integration of chaotic ODEs,
even long-time computation of the finite-time Lyapunov exponents
and the finite-time Lyapunov dimension
does not necessary lead to a relevant approximation
of the Lyapunov exponents and the Lyapunov dimension.

On the one hand, computational errors (caused by a finite precision
arithmetic and numerical integration of differential equations)
and sensitivity to initial data allow one to get a more complete visualization
of chaotic attractor (pseudo-attractor) by one pseudo-trajectory
computed for a sufficiently large time interval.
On the other hand, there arises a question of the reliability of calculating
the trajectory itself and its various characteristics,
such as finite-time Lyapunov exponents (FTLEs) and finite-time Lyapunov dimension (FTLD),
over a long time interval.
In~\cite{KehletL-2013} for the Lorenz system~\eqref{sys:lorenz}
the time interval for reliable
computation with $16$ significant digits and error $10^{-4}$ is estimated
as $[0, 36]$, with error $10^{-8}$ is estimated as $[0, 26]$, and reliable
computation for a longer time interval, e.g. $[0, 10000]$ in \cite{LiaoW-2014},
is a challenging task.

\begin{figure}[h!]
 \centering
 \subfloat[]{
    \label{fig:rossler:upo1:attr}
    \includegraphics[width=0.3\textwidth]{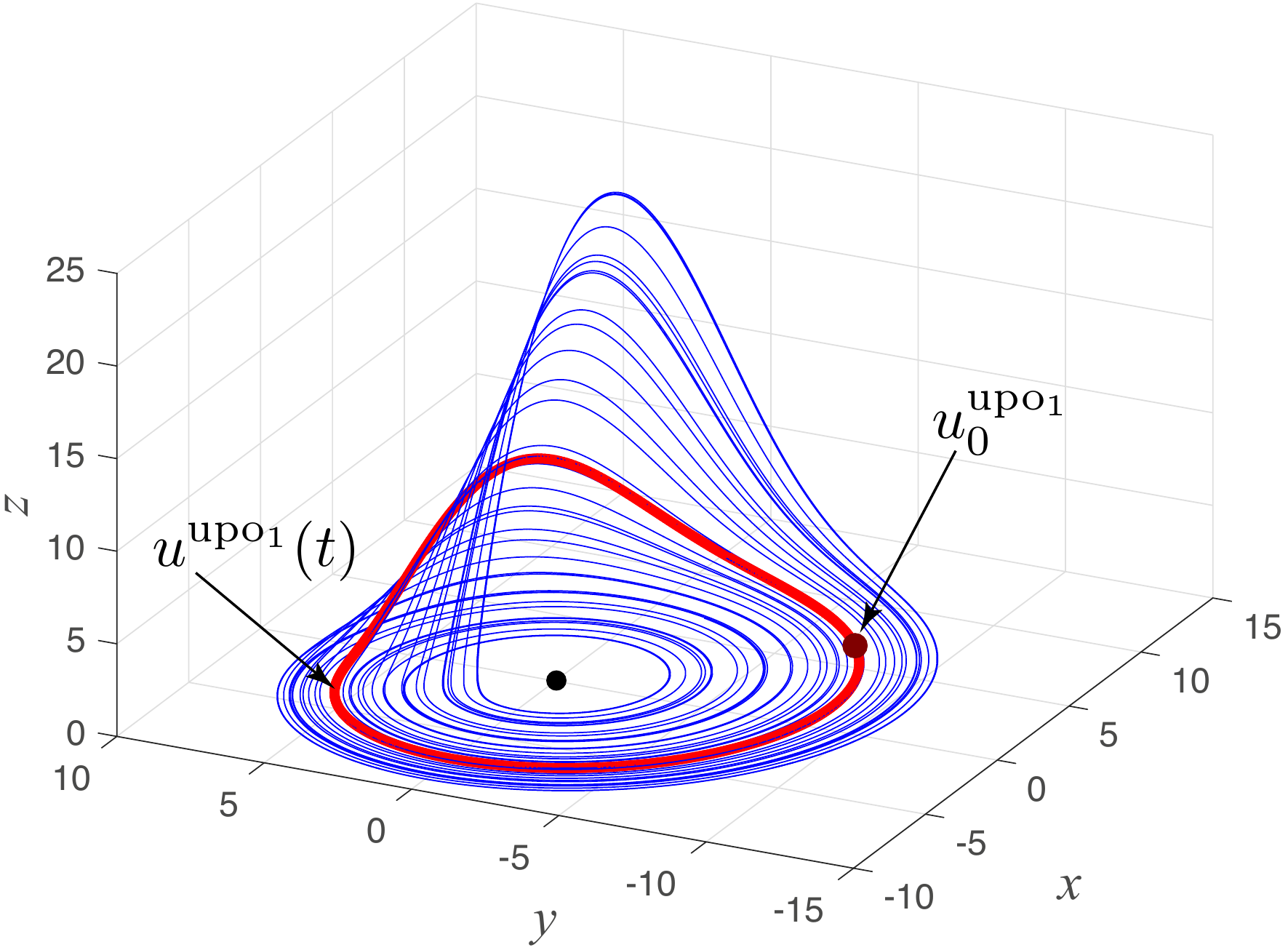}
  }
 \subfloat[]{
    \label{fig:rossler:upo1:LE1}
    \includegraphics[width=0.3\textwidth]{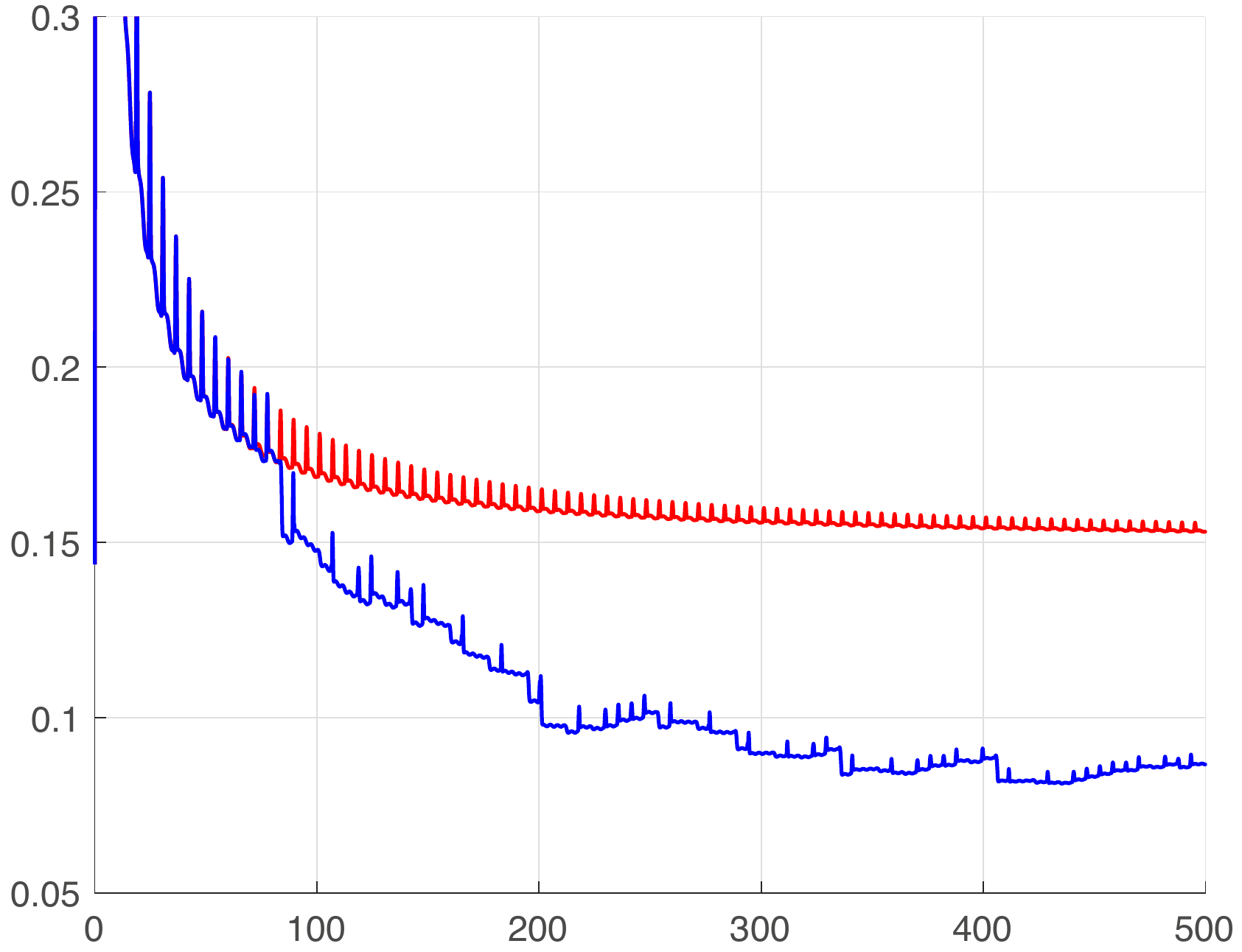}
  }
  \subfloat[]{
    \label{fig:rossler:upo1:LD}
    \includegraphics[width=0.3\textwidth]{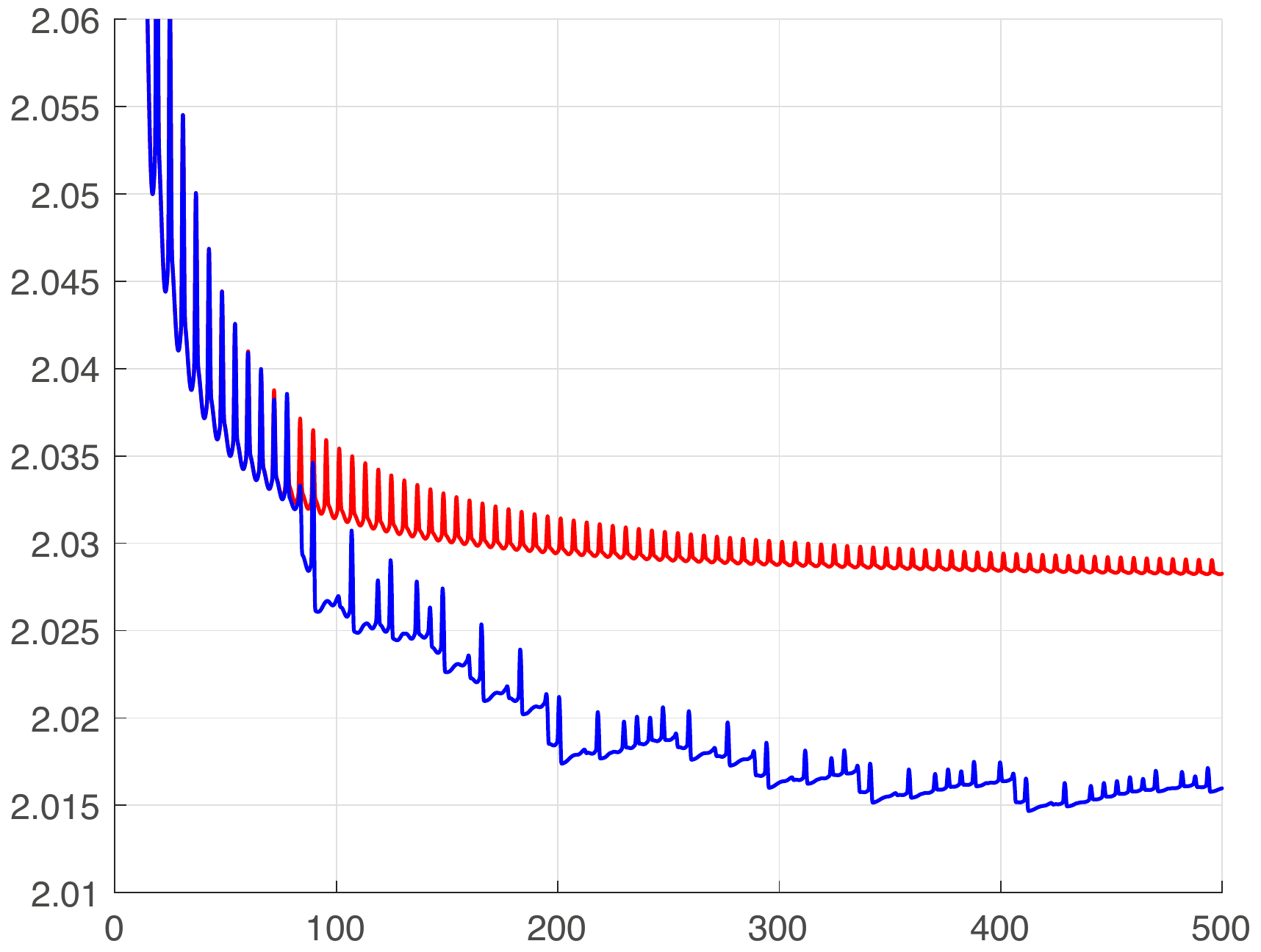}
  }
\caption{
Numerical computations of trajectory (a),
largest finite-time Lyapunov exponent $\LEs_1(t,u^{\rm upo_1}_0)$ (b),
and finite-time local Lyapunov dimension $\dim_{\rm L}(t, u^{\rm upo_1}_0)$ (c)
with (red) and without (blue) application of Pyragas’ correction control
on the time interval $t \in [0,500]$.
Initial point $u^{\rm upo_1}_0 = (6.491, -7.0078, 0.1155)$ (dark red)
is chosen at the computed UPO $u^{\rm upo_1}$.
}
\label{fig:rossler:upo1}
\end{figure}

For two different long-time pseudo-trajectories
$\tilde u(t,u^1_0)$ and $\tilde u(t,u^2_0)$ visualizing the same attractor,
the corresponding FTLEs can be, within the specified accuracy,
similar due to averaging over time and similar sets of points
$\{\tilde u(t,u^1_0)\}_{t \geq0}$ and $\{\tilde u(t,u^2_0)\}_{t \geq0}$.
At the same time, one of the corresponding real trajectories $u(t,u^{1,2}_0)$
may correspond to an unstable periodic orbit (UPO) which is embedded in the attractor
and does not allow one to visualize it.
The limitations of the possibilities of numerical integration procedures
demonstrates~\cite{KuznetsovM-2018-arXiv} an example of
the R\"{o}ssler system~\cite{Rossler-1976}:
\begin{equation}\label{sys:rossler}
\begin{cases}
 \dot{x} = - y - z, \\
 \dot{y} = x + 0.2 y, \\
 \dot{z} = 0.2 - 5.7 z + x z.
\end{cases}
\end{equation}
For system \eqref{sys:rossler} it is possible to stabilize
an unstable periodic orbit (UPO) $u^{\rm upo_1}$ with period $\tau \approx 5.8811$
embedded into attractor.
The corresponding computations by the standard MATLAB numerical integration procedure
with and without application of the Pyragas' correction control \cite{Pyragas-1992}
(see also \cite{KuznetsovLS-2015-IFAC})
of the largest FTLE, $\LEs_1(t,u^{\rm upo_1}_0)$, and
FTLD $\dim_{\rm L}(t, u^{\rm upo_1}_0)$ along a trajectory with
initial data $u^{\rm upo_1}_0 \in u^{\rm upo_1}$
over the time interval $[0, 500]$
give us the following results.
On the initial part of the time interval,
one can indicate the coincidence of these values
with a sufficiently high accuracy.
For the UPO and for the unstabilized trajectory
$\LEs_1(t,u^{\rm upo_1}_0)$
coincide up to the 5th decimal place inclusive on the interval $[0,30.4]$,
up to the 4th decimal place inclusive on the interval $[0,53.8]$,
up to the 3rd decimal place inclusive on the interval $[0,71.5]$.
After $t > 71.5$ the difference in values becomes significant and the
corresponding graphics diverge in such a way that the part of the graph
corresponding to the unstabilized trajectory
is lower than the part of the graph corresponding to the UPO
(see Fig.~\ref{fig:rossler:upo1:LE1}).
Thus, the application of the Pyragas' procedure makes it possible to
compensate round-off errors and to trace the UPO numerically\footnote{
  There are well-known cases when the accumulation of errors in
  the computer representation of real data led to catastrophes (see, e.g. \cite{Skeel-1992})
}.
Note that other UPOs could be revealed, e.g., by various evolutionary algorithms~\cite{Zelinka-2015}.

\section{Analytical localization of attractors via Lyapunov functions}
In order to simplify the numerical search for attractors,
one can apply an analytical localization approach related to the dissipativity
in the sense of Levinson~\cite{LeonovKM-2015-EPJST}.
Also this approach may help to obtain an exact formula or to estimate
Lyapunov dimension in the entire phase space.
As an example of the effectiveness of such approach,
we consider the localization of attractors
in the Vallis system describing
the El Ni\~{n}o-Southern Oscillation (ENSO) phenomenon of irregular,
anomalous, Christmas time warming of the coastal waters
of Peru and Ecuador about every 3–6 years that
affected weather on a global scale.
A low-order model for the ENSO phenomenon
was suggested by G.K. Vallis~\cite{Vallis-1986}
and has the following form:
\begin{equation}\label{sys:vallis}
  \begin{cases}
    \dot{x} = B y - C (x + p), \\
    \dot{y} = x z - y, \\
    \dot{z} = -x y - z + 1,
  \end{cases}
\end{equation}
with parameters $B, C > 0$, and real~$p$.
If $p = 0$, then system \eqref{sys:vallis}
can be transformed to the Lorenz system \eqref{sys:lorenz}.
The number of equilibria in system \eqref{sys:lorenz}
depends on the sign of the discriminant
\begin{equation*}\label{eq:discrimCube}
    D = \tfrac{Q^2}{4}+\tfrac{P^3}{27}, \quad
    Q = \tfrac{p}{3}\left(\tfrac{2 p^2}{9} + \tfrac{B}{C} + 2\right), \quad
    P = - \left(\tfrac{p^2}{3} + \tfrac{B}{C} - 1\right)
\end{equation*}
of the cubic equation
\begin{equation}\label{eq:xCoord}
    x^3 + p \,x^2 + \left(1-\tfrac{B}{C}\right) x + p = 0.
\end{equation}
If $B < C$, then for any real $p$ system~\eqref{sys:vallis} has only one equilibrium,
Otherwise, if $B > C$, and also
$p \in (-p^*,p^*), \quad p^* =
        \sqrt{\left(\tfrac{B^2}{8C^2} - \tfrac{5B}{2C} - 1 \right)
        + \tfrac{1}{8}\sqrt{\tfrac{B}{C}\big(\tfrac{B}{C}+8\big)^3}}$
we get $D < 0$ and system \eqref{sys:vallis} has three equilibria
$O_j = (x_j, y_j, z_j)$, $j = 1,2,3$,
where $x_j$ are the solutions of \eqref{eq:xCoord}, and $y_j = \frac{C (x_j + p)}{B}$,
$z_j = \frac{C (x_j + p)}{B x_j}$.
One can express $x_j$ as follows
\begin{equation}\label{equil:compl}
x_j = - \frac{p}{3} + \xi^{j-1} \sqrt[3]{-\tfrac{Q}{2}+\sqrt{D}} \\
+ \xi^{2(j-1)} \sqrt[3]{-\tfrac{Q}{2}-\sqrt{D}}, \quad
j = 1,2,3, \quad \text{where} \quad \xi = -\tfrac{1}{2}+\tfrac{\sqrt{3}}{2} i.
\end{equation}

Using the direct Lyapunov method, we can prove
the dissipativity of system \eqref{sys:lorenz} for $C \geq \frac{1}{2}$ and
obtain the following ellipsoidal absorbing set:
\begin{equation}\label{absorb_set2}
\mathcal{B}\!=\!\left\{\!(x,y,z)\!\in\!\mathbb{R}^3 \,\big|\,
\tfrac{1}{B} x^2\!+\!y^2\!+\!(z\!+\!1)^2\!\leq\!4\!+\!\tfrac{(Cp)^2}{(2C-1)B}\!\right\}\!.
\end{equation}

Let $u = (x,\, y, \, z) \in U = \mathbb{R}^3$,
and the dynamical system $\{\varphi^t_{\rm V}\}_{t\geq0}$,
is generated by the Vallis system \eqref{sys:vallis}
with positive parameters $B$, $C$ and real parameter $p$,
and $\mathcal{A}_{\rm V} \subset \mathbb{R}^3$ is
a nonempty closed bounded set, which is invariant
with respect to the dynamical system $\{\varphi_{\rm V}^t\}_{t \geq 0}$.
i.e. $\varphi^t_{\rm V}(\mathcal{A}_{\rm V}) = \mathcal{A}_{\rm V}$ for all $t \geq 0$.
Using an effective analytical approach,
proposed by Leonov~\cite{Leonov-1991-Vest,Kuznetsov-2016-PLA},
which is based on a combination of the Douady-Oesterl\'{e} approach with the direct Lyapunov method
we obtain the upper estimate of the Lyapunov dimension for the global attractor in system \eqref{sys:vallis}.
\begin{theorem}\label{thm:main}
For the Vallis dynamical system $\{\varphi^t_{\rm V}\}_{t \geq 0}$,
generated by system \eqref{sys:vallis} with $B,C > 0$, and real $p$
we have the following estimate for the Lyapunov dimension of
it global B-attractor
\begin{equation}\label{thm:estim}
  \dim_{\rm L} \mathcal{A}_{\rm V}
  \leq 3 - \tfrac{2(C + 2)}{C + 1 + \sqrt{(C - 1)^2 + \tfrac{25}{4} B}}.
\end{equation}
\end{theorem}

Let us compare the obtained estimate \eqref{thm:estim} with the values
of local Lyapunov dimension at the equilibria $O_1$, $O_2$, $O_3$.
E.g., for $B = 102$, $C = 3$, $p = 0.83$ we get the following local Lyapunov dimensions
\[
\dim_{\rm L} O_1 = 2.58834, \
\dim_{\rm L} O_2 = 2.00368, \
\dim_{\rm L} O_3 = 2.07175.
\]
and the corresponding estimate \eqref{thm:estim}
is in accordance with the values of local Lyapunov dimensions
\[
\dim_{\rm L} O_2 < \dim_{\rm L} O_3 < \dim_{\rm L} O_1 \leq \dim_{\rm L}\mathcal{A} \leq 2.65903.
\]

\section*{Acknowledgement}\label{sec:acknowledgement}
The work is supported by the Russian Science Foundation project (14-21-00041).
This article is based on the lecture at the VII International Conference
''Problems of Mathematical Physics and Mathematical Modelling''
(NRNU MEPhI, Moscow, 2018) and is accepted for publication
in Journal of Physics: Conference Series.

\section*{References}
\bibliographystyle{iopart-num}

\providecommand{\newblock}{}

\end{document}